\documentclass[manuscript]{acmart}

\AtBeginDocument{%
  }

\acmISBN{978-1-4503-XXXX-X/18/06}

\begin{document}

%%
%% The "title" command has an optional parameter,
%% allowing the author to define a "short title" to be used in page headers.
\title{Algorithmic Autonomy in Data-Driven AI}

%%
%% The "author" command and its associated commands are used to define
%% the authors and their affiliations.
%% Of note is the shared affiliation of the first two authors, and the
%% "authornote" and "authornotemark" commands
%% used to denote shared contribution to the research.

\author{Ge Wang}
\affiliation{%
  \institution{Stanford University}
  \country{USA}}

\author{Roy Pea}
\affiliation{%
  \institution{Stanford University}
  \country{USA}}

%%
%% By default, the full list of authors will be used in the page
%% headers. Often, this list is too long, and will overlap
%% other information printed in the page headers. This command allows
%% the author to define a more concise list
%% of authors' names for this purpose.
% \renewcommand{\shortauthors}{Wang et al.}

%%
%% The abstract is a short summary of the work to be presented in the
%% article.
\begin{abstract}
  In societies increasingly entangled with algorithms, our choices are constantly influenced and shaped by automated systems. This convergence highlights significant concerns for individual autonomy in the age of data-driven AI. It leads to pressing issues such as data-driven segregation, gaps in accountability for algorithmic decisions, and the infringement on essential human rights and values. Through this article, we introduce and explore the concept of \textit{algorithmic autonomy}, examining what it means for individuals to have autonomy in the face of the pervasive impact of algorithms on our societies. We begin by outlining the data-driven characteristics of AI and its role in diminishing personal autonomy. We then explore the notion of algorithmic autonomy, drawing on existing research. Finally, we address important considerations, highlighting current challenges and directions for future research.
\end{abstract}

\maketitle

\section{Main}

AI-based platforms, by integrating artificial intelligence, are transforming how we navigate digital ecosystems. These platforms analyze vast data sets, enabling personalized content recommendations in areas like social media, music, gaming, and shopping, enhancing learning through intelligent tutoring systems, filtering harmful content for safer online environments, and making digital interactions more human-like through social robots. Despite these benefits, the algorithmic nature of AI introduces new risks that are often insufficiently addressed. \textit{Datafication}—the recording, tracking, aggregating, analyzing, and monetizing of user data—is evident in how AI-based platforms manage people’s data and activities~\cite{mascheroni2020datafied}. As AI becomes more integrated into smart devices, datafication, surveillance, and behavioral engineering intensify. Algorithms assess personal characteristics, making decisions that shape lives, turning smart devices into active agents in our daily experiences. This can significantly undermine people's autonomy, affecting their content consumption, behavior, and self-regulation. Yet, efforts to enhance autonomy are often overlooked in AI development, leaving individuals ill-equipped to make informed choices. This article aims to introduce the concept of \textbf{algorithmic autonomy}, referring to individuals' capacity to make independent decisions within algorithmic systems. We explore how AI's data-driven nature erodes autonomy and unpack the concept of algorithmic autonomy' based on existing literature. We then address key challenges and propose directions for future research.

\section{Data-Driven AI as Problematic Practices for Users' Autonomy}

\textit{Data-Driven AI}, a subset of artificial intelligence where algorithms learn and make decisions by analyzing large datasets, is becoming the mainstream approach in today's AI landscape. Unlike traditional model-driven methods that rely on predefined rules and logic data-driven AI systems use advanced algorithms to identify patterns within data, allowing machines to perform tasks and improve over time without explicit programming for each instruction. This capability enhances their ability to make predictions, recognize images, understand language, and more~\cite{ntoutsi2020bias}. AI platforms' effectiveness is rooted in the socio-technical system enabling datafication, the process of recording, tracking, aggregating, and analyzing user data. By leveraging this data, AI platforms can influence and potentially manipulate users' online behavior and engagement~\cite{mascheroni2020datafied}. Platforms make data inferences about users, predicting aspects like performance, economic situation, health, preferences, and behavior~\cite{mascheroni2020datafied}. As algorithms exploit user data, AI platforms gain power to fine-tune and influence users' beliefs, interests, and actions, facilitating micro-targeting and fostering reliance on these platforms. These practices, increasingly common across online platforms, have significant implications for how individuals perceive and interact with the world~\cite{buchi2020chilling}.

As explained above, data-driven AI is increasingly playing a pivotal role in decision-making for individuals. Even as these technologies offer unprecedented opportunities and advancements, they introduce new forms of risks, particularly by undermining personal autonomy when algorithmic decisions about individuals are made exploiting their data. Here, we articulate three ways in which people may lose autonomy within the realm of data-driven AI: (1) data-driven segregation, (2) responsibility voids, and (3) infringing upon human rights and values.

\subsection{Data-driven segregation}

Data-driven AI systems employed by major online platforms like Google, Yahoo, and Meta (including Facebook and Instagram) rigorously collect and analyze vast arrays of data—including demographic details, interests, and attitudes—to predict future actions, characteristics, and preferences of individuals or groups~\cite{rao2015they}. An example is Facebook analyzing user data to create interest classifications to boost advertiser sales and engage users on their news feed~\cite{thorson2021algorithmic}. Although these practices might initially appear benign, ostensibly aimed at enhancing content personalization by aligning with users' predicted preferences, they inadvertently create the foundation for data-driven segregation. This segregation stems from algorithmic decisions which, deliberately or not, exacerbate societal divides by relying on biased datasets, flawed algorithms, or discriminatory practices within their technological infrastructure. This issue is deeply rooted in the very foundation of modern AI systems, where the data-driven approach not only mirrors but also amplifies existing societal biases encoded within the data. By personalizing user experiences, these algorithms may mercurially reinscribe extant  worldviews or biases and deepen divides. Examples include online advertising, such as selective targeting of specific demographic groups for job advertisements, educational opportunities, or housing listings, systematically excluding others and perpetuating societal inequalities in an internet web services version of red-lining. Predictive policing exacerbates this issue by directing disproportionate surveillance and enforcement efforts towards particular neighborhoods, negatively impacting marginalized communities the most~\cite{o2019challenging}. In cases like these, data-driven AI contributes to societal divisions and restricts individuals' ability to influence their own choices, lives, and broader societal outcomes.

\subsection{``The problem of no hands''}

The data-driven nature of AI poses significant risks to user autonomy through responsibility voids. Similar to the ``problem of many hands'' observed in collective decision-making~\cite{van2015problem}, complexity arises because decisions are often based on aggregated data from diverse sources, leading to a diffusion of responsibility. Meanwhile, responsibility voids also occur as data-driven systems aggregate and analyze information to make decisions without direct human intervention, epitomizing the ``problem of no hands''. Unlike traditional decision-making processes where accountability is more straightforward, data-driven AI obscures accountability. The algorithms themselves are designed by humans but operate and evolve in ways that can be difficult to predict or understand fully, even by their creators. Additionally, the collective nature of the data used, sourced from numerous individuals, adds another layer of complexity to pinpointing responsibility for these decisions. When actions are based on this intricate input, it becomes difficult to ascertain who should be accountable for mistakes, biases, or negative outcomes, whether it’s the algorithm's designers, the data providers, or the system operators. Therefore, the responsibility voids associated with data-driven AI manifest its core features: the opacity of its decision-making processes~\cite{lepri2018fair}. This opacity undermines human users' ability to understand, question, or seek redress for AI decisions that are consequential for them, highlighting a critical governance and ethical challenge that directly constrains their ability to influence and control their interactions with AI-driven environments. Recognizing these challenges, there has been a growing discourse on research and policies that have been developed for algorithmic accountability, including seminal works and guidelines from various authorities; however, new challenges arise in the translation of these principles into practice, including proven methods available as well as lack of robust legal and professional codes~\cite{wang2024challenges}.

\subsection{Infringes upon human rights and values}

By drawing inferences about individuals' lives, data-driven technologies extend beyond mere concerns of data protection by representing a tangible threat to human autonomy, propelled by increasingly advanced surveillance techniques. Scholars have extensively outlined how social media's grip over data, attention, and behavior significantly affects autonomy~\cite{sahebi2022social}. Companies profit from user data without fair compensation, while platforms may claim that free access to their services justifies data exchange. Beyond mere data exploitation, studies find that social media algorithms critically shape users' beliefs, interests, and actions, including political dialogue and personal values, often creating echo chambers or personalising content in ways that can promote radicalisation. This is particularly concerning when we center more vulnerable populations. For instance, studies indicate that Facebook exposed a vast majority of young adults in the UK to alcohol marketing monthly, including those underage~\cite{winpenny2014exposure}. Research underscores how these platforms can exploit peoples' insecurities by micro-targeting them with advertisements designed to superficially boost self-esteem~\cite{susser2019technology}. Moreover, the addictive nature of social media, driven by intermittent rewards, the fear of missing out, and a lack of natural stopping cues, further undermines autonomy. These practices fundamentally encroach individual autonomy and spotlight the critical need to address the ethical implications of data-driven technologies, underscoring a pressing concern for human rights and values.

\section{Introducing `Algorithmic Autonomy'}

As data-driven AI increasingly undermines people's autonomy, we emphasize the urgent need to restore control to individuals in their interactions with such systems. With this in mind, we introduce the concept of \textbf{algorithmic autonomy}. Our goal is to clarify this concept amidst the complexities of AI's \textbf{data-driven} decision-making and its profound user implications. To do so, we have reviewed the literature to explore current understandings of 'algorithmic' and 'autonomy.' By proposing a preliminary definition, we aim to identify and discuss key themes relevant to algorithmic autonomy, rather than creating a rigid or exhaustive framework, while speculating on the future of AI and its relationship with individuals.

\subsection{Unpacking `Algorithmic' in Algorithmic Autonomy.} 

To refine the scope of 'algorithmic' in the context of data-driven AI, we adapt Solove's privacy taxonomy~\cite{solove2002conceptualizing} and discussions on datafication~\cite{cukier2013rise,zuboff2019age,mejias2019datafication} to emphasize its algorithmic aspects. Solove's taxonomy, which categorizes data concerns into information collection, processing, dissemination, and invasion~\cite{solove2002conceptualizing}, provides a framework for understanding algorithmic processes. These include algorithmic collection (observation and recording by algorithms), algorithmic processing (data manipulation by algorithms), algorithmic dissemination (distribution of algorithmic conclusions), and algorithmic invasion (intrusions into personal or decision-making processes). Additionally, we consider datafication—the conversion of interactions into quantifiable data—highlighting the algorithmic infrastructure behind data collection, processing, and value-creation mechanisms like analysis, surveillance, and monetization, often controlled by large corporations and states. From these frameworks, we identify three key algorithmic elements relevant to algorithmic autonomy in data-driven AI.

\begin{itemize}
    \item Algorithmic Data Collection and Dissemination: Solove describes this process as "the watching, listening to, or recording of an individual's activities"~\cite{solove2002conceptualizing}, emphasizing its role in the digital realm. This element involves both the systematic collection of user data for algorithmic analysis and the subsequent sharing of that data. Algorithms are designed to extract data from users, analyze interactions and behaviors to build detailed profiles, and facilitate targeted data sharing with internal teams, partner organizations, third-party vendors, or advertisers. This dual role in data extraction and dissemination shapes the strategies and pathways through which data is circulated within and beyond digital platforms.
    \item Algorithmic Processing: This stage involves the sophisticated analysis and strategic application of collected data by algorithms. On digital platforms, it refers to how algorithms process user data to customize services and content, aligning with each user's unique preferences and behaviors. This process not only showcases the interplay between algorithms and user engagement but also underscores the algorithms' ability to shape user experiences. By using varied models, platforms offer highly personalized experiences, dynamically guiding user interactions and molding their digital environments through predictive analytics and continuous personalization.
    \item Algorithmic Inference: This aspect involves algorithms' advanced ability to evaluate and predict intimate personal aspects, such as job performance, economic status, or health. Building on Solove's concepts of data dissemination and invasion~\cite{solove2002conceptualizing}, algorithmic inference highlights the profound impact of algorithms, which not only process data but also learn from it to uncover deeply personal insights. This goes beyond simple data management, involving the drawing of nuanced conclusions, potentially revealing sensitive information like sexual orientation~\cite{wang2018deep}. While these inferences can enhance personalization, they also raise significant privacy concerns by exposing private aspects of an individual's life that they may not have intended to share.

\end{itemize}

\begin{figure*}[h]
\begin{center}
\includegraphics[width=0.65\textwidth]{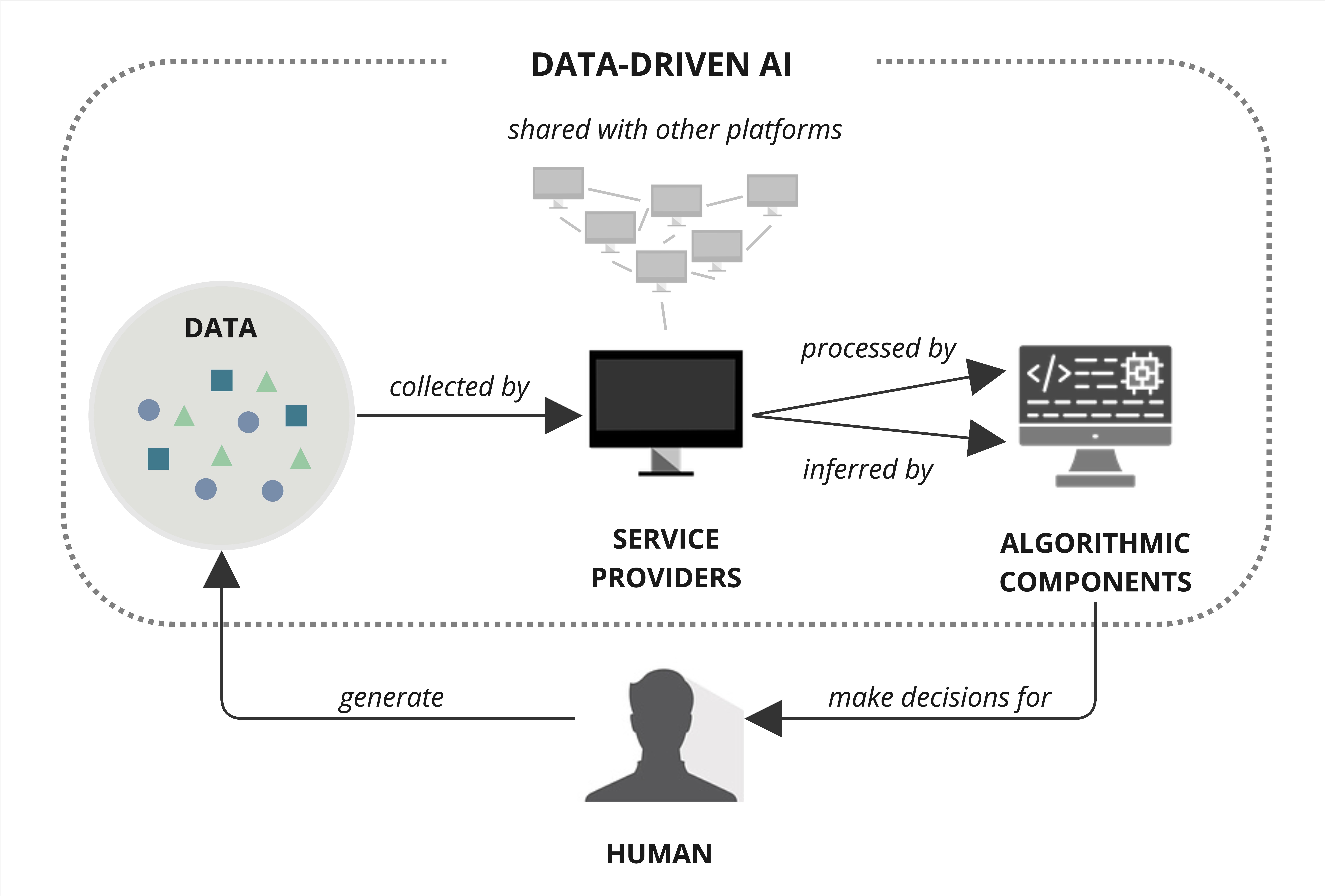}
\caption{A simplified model of interaction between humans and data-driven AI.}
\label{fig:concept}
\end{center}
\end{figure*}

\subsection{Unpacking `Autonomy' in Algorithmic Autonomy.} 

Autonomy is broadly defined as the state of being independent and self-governing, capable of making and acting on one’s own choices without external interference~\cite{spear2004autonomy}. Philosophers like Hurst Hannum, Ruth Lapidoth, Markku Suksi, and Yash Ghai have explored various dimensions of autonomy, from personal and behavioral to functional, cultural, and legislative~\cite{tkacik2008characteristics}. Our focus is on \textit{personal autonomy}, the ability to set goals based on personal values, closely linked to \textit{behavioral autonomy} (acting independently toward these goals) and \textit{functional autonomy}, which includes motivations tied to lifestyle or self-esteem and daily living activities. In self-determination theory, autonomy involves goal-setting, initiative, and effective self-regulation, allowing individuals to proactively direct their actions toward fulfilling their needs. It's important to distinguish between autonomy and agency. While often used interchangeably, they differ: agency is the capacity to act and make decisions, whereas autonomy is the freedom to make those decisions based on personal values, free from external influence~\cite{cummins2014agency}. This distinction is crucial in the context of data-driven AI, where external manipulation and behavioral engineering are significant concerns. An interesting metaphor in AI literature compares an agent to a coffee cup containing liquid—it has a purpose (agency) but not autonomy, as it cannot generate its own goals~\cite{luck1995formal}. We advocate for autonomy as it represents deeper self-direction and self-determination, reflecting not just the capacity to act, but the empowerment to act in ways that align with one’s beliefs, values, and desires. Building on these ideas, we identify three key components of autonomy:

\begin{itemize}
    \item \textit{Cognitive Autonomy} lays the foundational framework, empowering individuals with the ability to think independently. This involves critically assessing information, forming personal beliefs, and voicing opinions based on a self-governed process of knowledge acquisition and evaluation. It's the bedrock that enables one to discern their values and make informed decisions, which naturally leads to the next form.
    \item \textit{Behavioral Autonomy} builds upon this cognitive base, translating thought into action. It reflects an individual's capacity to act independently and make decisions aligning with their personal judgment and values. This form of autonomy is about the execution of one's free will, encompassing the ability to self-regulate, take responsibility for one's actions, and live according to one's choices and convictions. It represents the practical application of cognitive autonomy in one's daily life and interactions.
    \item \textit{Functional Autonomy} represents the evolution of motivations and behaviors to become self-driven, illuminating how actions or goals, initially perhaps externally motivated, become intrinsic parts of one's identity and purpose. This advanced form of autonomy shows full maturation of an individual, as cognitive and behavioral autonomies converge into a self-sustaining cycle of motivation and action. It signifies the stage where individuals no longer act out of external compulsion or immediate reward but are driven by deeply embedded, self-aligned motives.
\end{itemize}

Together, these forms of autonomy weave a narrative of personal growth, from the development of independent thought (Cognitive Autonomy) through to the independent execution of these thoughts in action (Behavioral Autonomy), culminating in a developmental state in which actions are inherently self-motivated and self-rewarding (Functional Autonomy). Building on the aforementioned concepts of ``algorithmic'' and ``autonomy'', one has \textbf{algorithmic autonomy} according to this definition:

\vspace{0.8cm} 

\noindent\fbox{%
    \parbox{\textwidth}{%
        \begin{definition}[Algorithmic Autonomy]
            Algorithmic Autonomy embodies the ability of individuals to understand and shape their interactions with data-driven AI systems. This competency requires employing cognitive autonomy for critical analysis, exercising behavioral autonomy for independent action, and achieving functional autonomy to align interactions with personal values. Specifically, it subsumes discerning the collection and gathering of one's own data, comprehending how algorithms process this information to tailor user experiences, and critically scrutinizing algorithmic inference to understand how AI predicts or influences one's personal behaviors and preferences.
        \end{definition}
    }%
}

\vspace{0.8cm}

\section{Challenges and Opportunities}

In our era of rapid advancements in data-driven AI and its societal impact, achieving algorithmic autonomy is crucial for human development. It empowers individuals by giving them control over their experiences and fostering responsibility for their decisions. However, attaining algorithmic autonomy is challenging and often overlooked in AI development literature or viewed through a different lens, such as user experience (e.g., the FATE principles~\cite{microsoft}). In this discussion, we explore the key challenges and opportunities in empowering individuals with algorithmic autonomy in the context of data-driven AI.

\subsection{Agency-fostering infrastructures}

To achieve true autonomy, it's essential to empower individuals to recognize and exercise their agency—the capacity to act independently and make informed choices, closely linked to \textit{behavioral autonomy}. This concept emphasizes not just making choices but actively following through on them. Studies often show that users feel a reduced sense of agency within data-driven AI systems~\cite{lukoff2021design, cornelio2022sense}. This sense of agency, the belief in one's ability to effect change, is crucial. For example, users with an agentic mindset towards social media—seeing it as within their control—experience less depression, anxiety, and stress~\cite{lee2022effects}. However, whether users genuinely control their data in AI-driven environments raises issues like the transparency paradox~\cite{bernstein2012transparency}. Despite privacy controls, a disconnect persists; an IBM study found that while 81\% of consumers worry about online data usage, many still consent to data collection without hesitation~\cite{ibmreport}.

Why, then, don't users' actions match their concerns? A significant missing point here is the lack of infrastructures for supporting users to realise their potential agency in data-driven AI. Data's intangible nature complicates matters further. Our data is not an asset which we deliberately hand off to someone; instead it is passively collected as we navigate online, easily eluding our attention and thus skirting potentially protective actions. This subtlety in data collection poses a formidable barrier for exerting control over our personal information, as the process largely remains hidden, revealing itself only through tangible outcomes like targeted marketing, free services, and customized digital experiences. At the heart of the problem is users' limited ability to dictate \textit{what data} are collected about them, compounded by a systemic lack of support for influencing how this data is processed and used. The challenge extends beyond individual capabilities to control their data; it's entrenched in the centralized architectures of current data-driven AI systems, which are fundamentally designed to dodge user agency in the process~\cite{kokolakis2017privacy}.

This structural gap highlights the need for new infrastructures in AI development which will empower agency by enabling users to actively participate in and significantly influence algorithmic decision-making. Opportunities for enhancement are evident throughout the spectrum of algorithmic processes. In data collection, a move towards ethical data governance is emerging~\cite{ewada}, driven by research aimed at decentralizing data control. Such efforts strive to increase individual control over personal data through mechanisms like collective access requests facilitated by NGOs and trade unions. For algorithmic processing, future research directions include human-AI collaboration~\cite{wang2020human} that invites users' critical input, offering them the tools to shape their digital experiences. Finally, concerns around algorithmic inference offers a significant opportunity for user empowerment by revealing to users how their data is profiled and granting them control over cross-platform tracking and profiling~\cite{10.1145/3544548.3580933}.

\subsection{Motivation-promoting mechanisms}

Autonomy, agency, and motivation are closely related yet distinct elements crucial for individual empowerment in the digital age. Autonomy involves self-governance and decision-making, while agency is the capacity for independent action. Motivation links the two, driving purposeful actions that align with personal values. Recent studies show that as users become accustomed to the algorithms shaping their online experiences, they often view these processes as mere enhancements to product quality. Zuboff et al.\cite{zuboff2019age} notes that users expect algorithms to deliver convenience without critically reflecting on the values they promote. Studies\cite{lepri2018fair,van2019know} suggest users are encouraged to trust these systems unconditionally, assuming they always serve their best interests. This blind trust can lead to manipulation, where users are subtly guided toward choices that align with system objectives rather than their own desires. For example, social network algorithms often prioritize polarizing content to boost engagement, subtly shaping user preferences~\cite{tu2022viral}. This normalization of algorithmic processes challenges users' understanding, even among those aware of these practices. The intangible nature of data manipulation makes it difficult for individuals to fully grasp how algorithms curate their experiences, leading to diminished motivation to actively manage interactions with data-driven AI systems, despite documented efforts by some users, such as on Twitter~\cite{burrell2019users}.

These aforementioned gaps underscore the need for new mechanisms for accelerating users' motivations towards algorithmic autonomy, a concept closely related to \textit{cognitive autonomy}. Awareness is undeniably the first crucial step toward achieving such autonomy, laying the groundwork for deeper critical reflection and the active exertion of control. While extensive existing research has focused on enhancing the transparency of algorithmic processes~\cite{rader2018explanations}, transparency alone is insufficient. It is imperative to move beyond mere transparency by empowering users with the recognition that they possess rights within a datafied society (e.g., to an explanation supporting understanding of algorithmic decision-making~\cite{kim2022right}). This awareness of rights is critical, as it underscores the significant implications of algorithmic processing on individual autonomy and privacy. Users must be encouraged to understand and exercise their rights to manage their digital footprints actively. Encouraging this level of engagement requires educational initiatives, clear policy communication, and the development of infrastructuring tools that make exercising these rights absolutely straightforward and effective.

\subsection{Data-driven AI as installations}

At the heart of our discussion is the challenge of enhancing individual autonomy within data-driven AI systems. A pivotal question arises: with the right motivation-enhancing mechanisms and agency-supporting infrastructures in place, can individuals truly exercise free will and shape their algorithmic experiences as they desire? An intriguing perspective comes from installation theory~\cite{lahlou2018installation}, which offers a framework for understanding how human systems are designed to both support and direct individual behavior. According to this theory, installations are structured environments—ranging from restaurants and shoe shops to voting booths and airports—that, despite our free will, nudge us towards predictable and standardized behaviors. These environments possess their own dynamics, subtly guiding, framing, and sometimes controlling our actions through a combination of spatial design, social interaction, and institutional rules. This results in a ``cultural reactor'' that produces predictable behavior, balancing empowerment and control through various feedback mechanisms~\cite{lahlou2015social}.

We argue that data-driven AI environments can be seen as installations, similar to how physical spaces like airports guide our actions through staged interactions—airline websites, check-in counters, security checks. AI systems guide us through digital landscapes, structuring our online experiences and influencing our behaviors in both enabling and regulatory ways. The choices presented to users—what content they see, which ads appear, how data are collected and used—are all preconfigured by algorithms reflecting societal norms and corporate objectives. As a result, individual autonomy in these systems is often limited to navigating a narrow set of predetermined options. This mirrors societal structures, indicating that autonomy in data-driven AI is not just about personal choice but also about the design of the broader digital ecosystem with its inherent biases. In this context, the distinction between \textit{epistemic} and \textit{deontic} modalities is crucial~\cite{pea1982world}. The epistemic dimension, concerning what users are enabled to do, reveals the system's capabilities, offering a semblance of autonomy through potential actions. However, the deontic dimension, focusing on what users are permitted or obligated to do, highlights the limitations and permissions dictated by the system's norms and values. This nuanced perspective suggests that while users \textit{can} navigate these digital spaces, their meaningful autonomy is further constrained by the deontic 'can do'—the permissions and obligations defined by the system—beyond just the capabilities of what they \textit{can do} in the epistemic sense.

This understanding opens the door to rich research avenues focused on dissecting the interaction dynamics between individuals and AI systems. Future studies should explore how individuals navigate their choices within algorithmically-curated environments and how these environments, in turn, shape user behavior. Potential research paths might include empirical experiments~\cite{lahlou2022multilayered} analyzing real-world user behavior in digital contexts and theoretical explorations into decision-making processes in varied algorithmic landscapes. Such research will be critical for designing interventions or support mechanisms aimed at modifying behavior or bolstering user autonomy in digital settings. By delving into these nuances, we can better design, regulate, and implement data-driven AI systems that respect and enhance our autonomy, rather than entrapping our agency.

\section{Outlook}

We live in societies deeply intertwined with algorithms, where our decisions are constantly influenced by automated systems. This raises significant concerns about individual autonomy in a data-driven AI era, leading to issues like data-driven segregation, accountability gaps in algorithmic decisions, and encroachments on fundamental human rights. In this article, we explore the concept of algorithmic autonomy, focusing on what it means for individuals to maintain self-governance amid the pervasive influence of algorithms. Achieving such autonomy is challenging but opens new research pathways. Pursuing algorithmic autonomy involves developing infrastructures that support user agency, allowing individuals to tailor their algorithmic interactions, and mechanisms to enhance motivation, encouraging users to assert their rights in a datafied society. We call for empirical studies and comprehensive frameworks to better understand how individuals interact with AI systems and how these environments shape user behavior. Recognizing the need for algorithmic autonomy, we invite your participation in advancing this vision through research and development.

%%
%% The next two lines define the bibliography style to be used, and
%% the bibliography file.
\bibliographystyle{ACM-Reference-Format}
\bibliography{refs}

\end{document}